\documentclass[useAMS,usenatbib]{mn2e}
\usepackage{psfig, epsf, epsfig}

\title[Intermediate-mass BH or foreground NS?]{X-ray study of HLX1:
intermediate-mass black hole or foreground neutron star?}
\author[R. Soria et al.]{Roberto Soria$^{1}$\thanks{E-mail:
roberto.soria@mssl.ucl.ac.uk},
Luca Zampieri$^{2}$, Silvia Zane$^{1}$, 
Kinwah Wu$^{1}$\\ 
$^{1}$Mullard Space Science Laboratory, University College London, Holmbury St Mary, Surrey RH5 6NT, UK\\
$^{2}$INAF, Osservatorio Astronomico di Padova, Vicolo dell'Osservatorio 5, I-35122 Padova, Italy}

\begin{document}

\date{Accepted 2010 Aug 17 ... Received 2010 Jun 14}

\pagerange{\pageref{firstpage}--\pageref{lastpage}} \pubyear{2010}

\maketitle

\label{firstpage}

\begin{abstract}
We re-assess the {\it XMM-Newton} and {\it Swift} observations of HLX1, 
to examine the evidence for its identification as an intermediate-mass black hole.
We show that the X-ray spectral and timing properties are equally consistent 
with an intermediate-mass black hole in a high state, or with 
a foreground neutron star with a luminosity 
$\sim$ a few $10^{32}$ erg s$^{-1}$ $\sim 10^{-6} L_{\rm Edd}$, 
located at a distance of $\approx 1.5$--$3$ kpc. 
Contrary to previously published results, we find that the X-ray spectral 
change between the two {\it XMM-Newton} observations of 2004 and 2008 (going from 
power-law dominated to thermal dominated) is not associated 
with a change in the X-ray luminosity. The thermal component becomes 
more dominant (and hotter) during the 2009 outburst seen by {\it Swift}, 
but in a way that is consistent with either scenario.
\end{abstract}

\begin{keywords}
accretion, accretion discs --- X-rays: binaries --- X-rays: individual: HLX1 --- stars: neutron --- black hole physics.
\end{keywords}

\section{Introduction}

Several theoretical arguments have been suggested for the formation 
of black holes (BHs) with masses $\sim 10^3$--$10^4 M_{\odot}$, 
straddling the gap between stellar and nuclear supermassive BHs. 
However, solid observational evidence of their existence remains 
lacking or disputed. The bright, point-like X-ray source 
2XMM J011028.1$-$460421 (henceforth, HLX1 for simplicity), 
discovered by \citet{far09}, has been proposed as the first 
unambiguous identification of an intermediate-mass BH \citep[see also][]{god09,web10}.
The (variable) X-ray emission from this source clearly indicates an accreting 
compact object rather than a star. It appears located inside or projected in front 
of the bulge/halo of the S0 galaxy  \citep{far09}, located at a distance 
of $\approx 91$ Mpc \citep{afo05}. If HLX1 does belong to that galaxy, it reached 
X-ray luminosities $\approx 10^{42}$ erg s$^{-1}$, implying a BH mass 
$> 1000 M_{\odot}$ from Eddington-limit arguments \citep{far09}. 
Its optical counterpart \citep{sor10} is a point-like source with 
$R \approx 24$ mag, which implies an X-ray/optical flux ratio 
$\sim 500$--$1000$. This is consistent with an X-ray binary, 
but rules out a background AGN, for which we would expect 
flux ratios $\sim 0.1$--$10$. The brightness and colour of the optical 
counterpart are consistent with either a massive globular cluster in ESO243$-$49 
(which may contain an accreting intermediate-mass BH) 
or a foreground M star in the Galactic Halo \citep{sor10}.
A residual emission line consistent with H$\alpha$ redshifted 
by the systemic velocity of ESO243$-$49 \citep{far10} seems 
to provide decisive support for the intermediate-mass BH 
interpretation. However, the issue is still hotly debated.

In this paper, we discuss the constraints 
to the nature of HLX1 provided by the {\it XMM-Newton} 
and {\it Swift} observations, and re-examine its X-ray luminosity 
and spectral properties. In particular, we want to determine whether 
X-ray flux and spectral information are already sufficient to rule out 
the possibility of a foreground neutron star (NS), weakly accreting 
from a low-mass donor star. If that is the case, the X-ray 
properties of HLX1 could be used in the future as a template 
to identify other intermediate-mass BHs.

\section{X-ray observations}

{\it XMM-Newton}'s European Photon Imaging Camera (EPIC) 
observed HLX1 on 2004 November 23 (ObsID 0204540201: 
serendipitously and $\approx 10\arcmin$ off-axis) and on 2008 November 28 
(ObsID 0560180901: target observation, on axis); 
see Table 1 for a summary of instrument modes and live times. 
Henceforth, we will refer to those observations as XMM1 and XMM2.
We downloaded the Observation Data Files from the public archive, 
and used the Science Analysis System ({\footnotesize{SAS}})
version 9.0.0 ({\it xmmsas\_20090615}) 
to process and filter the event files and extract spectra.
We checked that there were no background flares in either observation.
For XMM1 MOS, we extracted the source spectra from a circular region 
of radius $45\arcsec$; for XMM1 pn, we used a $25\arcsec \times 30\arcsec$ 
ellipse, to reduce overlapping with a chip gap and the row of pixels 
next to it.  For XMM2 MOS and pn, we used a circular region 
of radius $30\arcsec$, because the source is on-axis 
and has a narrower point-spread function. We defined suitable 
background regions to avoid chip gaps.
We selected single and double events (pattern $\le 4$ for the pn and 
pattern $\le 12$ for the MOS). 
After building response and ancillary response files with the {\footnotesize{SAS}} 
tasks {\it rmfgen} and {\it arfgen}, we used {\footnotesize{XSPEC}} \citep{arn96} 
Version 12 for spectral fitting.

We used the XMM2 data for timing analysis (XMM1 being too short, and affected 
by the chip gap problem). The science modes used for XMM2 give 
a time resolution of 5.7 ms for the pn, and 2.6 s for the MOS detectors. 
We used {\it xmmselect} to extract pn and MOS 
source and background lightcurves, selecting single and double events 
in the $0.2$--$12$ keV range.
We defined the same start and stop time for the pn and MOS lightcurves, 
so that they could be combined into a total EPIC lightcurve.
This is possible because pn and MOS have a similar exposure times $\approx 50$ ks, 
although the pn live time is only 71\% of the exposure time 
(small window mode). Background subtraction, together 
with corrections for various sorts of detector inefficiencies 
(vignetting, bad pixels, dead time, etc.) was performed with the  
{\footnotesize{SAS}} task {\it epiclccorr}.
We then used standard {\footnotesize{FTOOLS}} tasks for timing analysis.

\begin{table}
\begin{center}
\begin{tabular}{lrrr}
\hline
Date & Instrument  & Mode & Live time\\
\hline\\[-5pt]
2004 Nov 23 & pn & prime full (thin1) & 18.0 ks\\[3pt]
 & MOS & prime full (thin1) & 21.6 ks \\[3pt]
2008 Nov 28 & pn & prime small (thin1) & 35.3 ks \\[3pt]
 & MOS & prime full  (thin1) & 49.9 ks \\[3pt]
\hline
\end{tabular}
\end{center}
\caption{{\it XMM-Newton} observation log. }
\end{table}

In addition, HLX1 has been the target of over 40 {\it Swift} X-Ray Telescope (XRT) 
observations since 2008 October; see NASA's Heasarc data archive for a detailed logbook. 
We used the on-line XRT data product generator
\citep{eva07,eva09} to extract lightcurves and spectra (including 
background and ancillary response files); we selected grade 0-12 events. 
We downloaded the suitable spectral response file for single and double events 
in photon-counting mode from the latest Calibration Database (2009 December 1); 
it is the same response used by \citet{god09}. 
We grouped the {\it Swift} spectra into four bands, according 
to count rates (Section 3.3), and fitted the coadded spectra of each band 
with {\footnotesize{XSPEC}} Version 12.

\begin{figure}
\begin{center}
\psfig{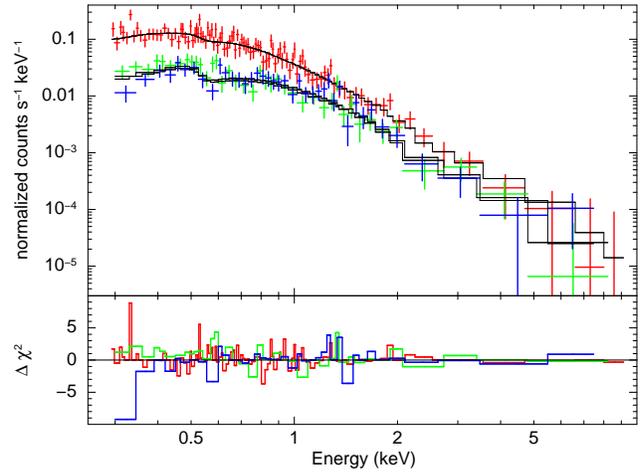}
\end{center}
\caption{{\it XMM-Newton}/EPIC spectra from the 2004 observation, 
simulataneously fitted with the neutron star atmosphere model {\it zamp} plus power-law. 
EPIC-pn datapoints and $\chi^2$ contributions are plotted in red; 
MOS1 data in green; MOS2 data in blue. See Table 2 for the best-fitting 
parameters.}
\label{f1}
\end{figure}

\begin{figure}
\begin{center}
\psfig{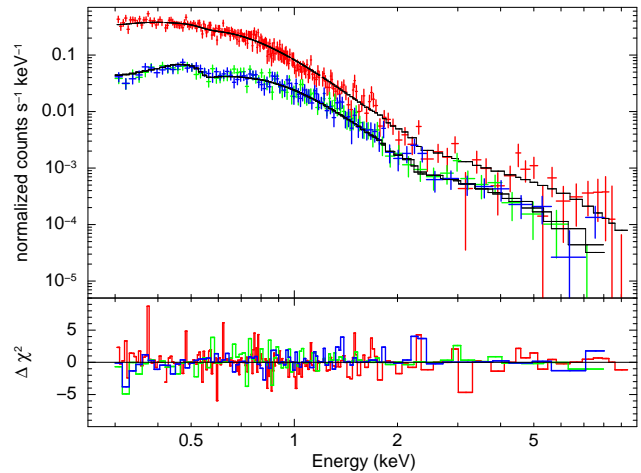}
\end{center}
\caption{As in Figure 1, for the 2008 {\it XMM-Newton}/EPIC spectra. 
See Table 3 for the best-fitting parameters.}
\label{f2}
\end{figure}

\section{X-ray spectral modelling}

\subsection{Choice of models}

We chose four spectral models based on the combination 
of a power-law with a soft thermal component, suitable 
to the high-accretion-state BH and/or 
the low-accretion-state NS scenarios.
The first model is power-law plus single-temperature 
blackbody, which gives the simplest phenomenological 
estimate of the soft component. The second model 
is power-law plus disk-blackbody, suitable to the BH scenario; 
the thermal disk component has a much broader spectral shape 
than the single-temperature blackbody. The models for thermal component 
in the third and fourth model are suitable to fit the emission from 
a weakly-magnetized NS hydrogen atmosphere in hydrostatic and 
local thermodynamical equilibrium: they are broader and harder 
than a simple blackbody at the same effective temperature, but slightly 
narrower than a standard disk-blackbody.

Blackbody ({\it bb} in {\footnotesize{XSPEC}}), disk-blackbody 
\citep[{\it diskbb}:][]{mak86}
and NS atmosphere \citep[{\it nsa}:][]{zav96} models are well known 
from the standard release of {\footnotesize{XSPEC}} and do not require 
additional explanations here. Our fourth model is 
the {\it zamp} model, which has been implemented in
{\footnotesize{XSPEC}} as an additive table \citep{cam97} 
using the spectra computed in \citep{zam95}.
The difference is that, while the {\it nsa} model computes the X-ray 
spectrum of a passively cooling NS, the {\it zamp} model was developed
specifically to reproduce the emission from non-magnetized
NSs accreting at very low rates
($10^{-7} \la L/L_{\rm Edd} \la 10^{-3}$), for example 
from the intesterllar medium, a molecular cloud, or a very low mass 
stellar donor (the last case may be applicable to HLX1).   
In fact, the X-ray spectra from {\it nsa} and {\it zamp} turn out to
be virtually indistinguishable at the signal-to-noise level of our data. 
The main reason for the similarity is that, at such low luminosities, 
the NS atmosphere develops smooth temperature and density 
gradients in the inner layers where free-free emission-absorption
dominates; those gradients are very similar in the two cases.
The free-free opacity is a function of frequency: as a result, 
the emerging higher-frequency photons are emitted 
in deeper (hotter) layers; the observed spectrum is a superposition 
of Planckians at different temperatures, with a broader plateau around 
the peak than a simple blackbody. In addition, the {\it zamp} model 
predicts a temperature inversion in the most external layers, due 
to accretion, but since this region is already optically thin 
to X-ray photons, it does not appreciably contribute 
to the observed X-ray spectrum.
The additional power-law component is observed in weakly accreting 
neutron stars at luminosities $\sim 10^{32}$ erg s$^{-1}$, 
although its origin is still unclear \citep{jon04}, perhaps associated 
to faint, residual magnetospheric activity.
In the high-state BH scenario, the power-law component comes 
from a hot Comptonizing medium that reprocesses part 
of the thermal disk photons.

For consistency, we have used the same NS mass $M = 1.4 M_{\odot}$ and 
true NS radius $R = 12.4$ km for both the {\it zamp} and {\it nsa} models.
The fitting parameter for the {\it nsa} model is the local (non-redshifted)  
effective temperature $T_{\rm eff}$; the effective temperature inferred  
by a distant observer is $T_{\rm eff} \left(1-2GM/R\right)^{0.5}$.
The fitting parameter for the {\it zamp} model is the total 
isotropic luminosity at infinity scaled to the Eddington luminosity, 
$L/L_{\rm Edd}$. This can be easily related to the effective temperature, 
because $L = 4\pi R^2 \sigma T^4_{\rm eff} \, \left(1-2GM/R\right)$, 
assuming that the NS is isotropically emitting from the whole surface 
(a scaling area factor can easily be introduced when this is not the case). 
The fitting parameter for the {\it bb} model is the colour temperature 
$T_{\rm bb}$ seen by the distant observer (that is, redshifted). Thus, when 
comparing the best-fitting temperature from those three models, 
we need to remember that 
$T_{\rm bb} = \gamma \left(1-2GM/R\right)^{0.5} T_{\rm eff} 
\approx 0.82 \gamma T_{\rm eff}$, 
where $\gamma$ is the hardening factor. 

\begin{figure}
\begin{center}
\psfig{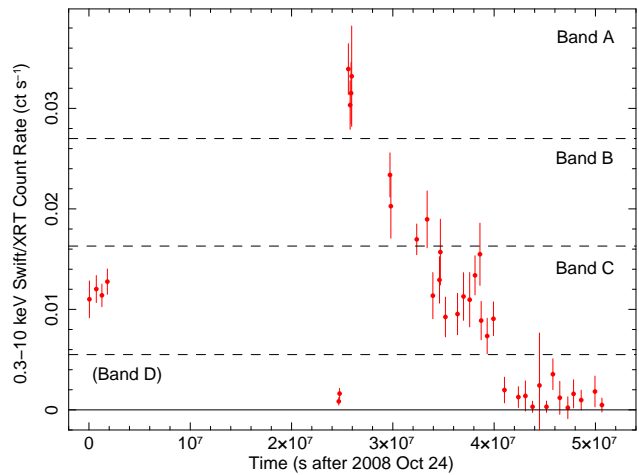}
\end{center}
\caption{{\it Swift}/XRT lightcurve from 2008 October to 2010 June, 
with our definition of count-rate bands. See Tables 4--6 for the 
best-fitting parameters to the coadded spectra from bands A, B and C.}
\label{f3}
\end{figure}

\begin{figure}
\begin{center}
\psfig{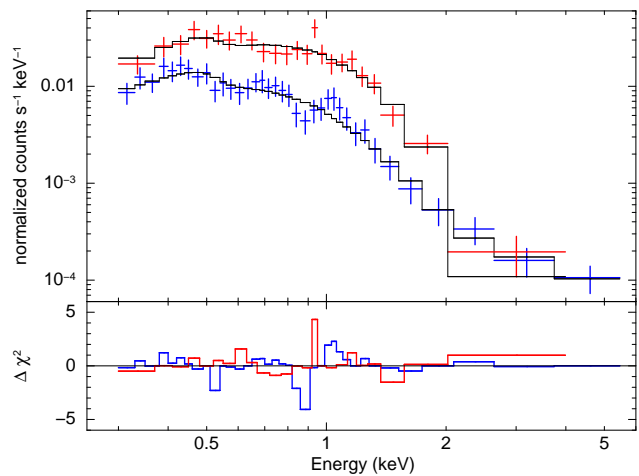}
\end{center}
\caption{{\it Swift}/XRT spectra from band A (red) and band C (blue) 
fitted with the neutron star atmosphere model {\it zamp} plus power-law.
The plot highlights the significant spectral variability between the two 
sets of (coadded) observations. See Figure 3 for our definition 
of {\it Swift}/XRT count-rate bands, and 
Tables 4--6 for the best-fitting parameters.}
\label{f4}
\end{figure}

\begin{table*}
\begin{center}
\begin{tabular}{lrrrr}\hline
\multicolumn{5}{c}{{\bf First XMM observation}} \\
\hline
 & wabs*(bb+po) & wabs*(diskbb+po) & wabs*(zamp+po) & wabs*(nsa+po)  \\
\hline
Parameter & \multicolumn{4}{c}{Value}\\
\hline\\[-5pt]
$N_{H,\rm int}$ & \ \ \ \ $4.1^{+2.8}_{-2.9} \times 10^{20}$ cm$^{-2}$ & \ \ \ \ 
        $4.3^{+3.6}_{-3.4} \times 10^{20}$ cm$^{-2}$ & \ \ \ \ 
        $3.6^{+3.5}_{-2.9} \times 10^{20}$ cm$^{-2}$ & \ \ \ \ 
        $3.1^{+5.0}_{-2.9} \times 10^{20}$ cm$^{-2}$ 
        \\[5pt]
$kT_{\rm bb}$  
        & $0.131^{+0.028}_{-0.027}$ keV &&&
        \\[5pt]
$N_{\rm bb}$  
        & $1.5^{+1.1}_{-0.9}  \times 10^{-6}$ &&&\\[5pt]
$kT_{\rm dbb}$ & 
        & $0.17^{+0.05}_{-0.04}$ keV &&
        \\[5pt]
$N_{\rm dbb}$ & 
        & $12.9^{+46.8}_{-10.1}$ &&\\[5pt]
$\log \left(L/L_{\rm Edd}\right)$ &&&  $-6.04^{+0.49}_{-0.45}$ &\\[5pt]
$N_{\rm zam}$ &&&  $2.0^{+1.3}_{-1.2} \times 10^{-5}$ & \\[5pt]
$kT_{\rm eff}$ & & & [$0.059^{+0.020}_{-0.013}$ keV]
        & $0.059^{+0.014}_{-0.007}$ keV
        \\[5pt]
$N_{\rm nsa}$ & & &
        & $1.8^{+32.3}_{-1.4} \times 10^{-7}$ 
        \\[5pt]
$\Gamma$ & $2.99^{+0.35}_{-0.35}$
         & $2.94^{+0.38}_{-0.43}$
         & $2.87^{+0.43}_{-0.40}$
         & $2.81^{+0.51}_{-0.28}$
         \\[5pt]
$N_{\rm po}$ & $6.4^{+2.0}_{-1.8} \times 10^{-5}$
         & $6.1^{+2.3}_{-2.2} \times 10^{-5}$
         & $5.4^{+2.4}_{-2.0} \times 10^{-5}$
         & $5.0^{+3.2}_{-1.9} \times 10^{-5}$
         \\[5pt]
\hline\\[-5pt]
$f_{0.3-10}$ & \ \ \ \ $2.7^{+0.3}_{-0.3} \times 10^{-13}$ 
        & $2.7^{+0.3}_{-0.6} \times 10^{-13}$
        & $2.7^{+0.4}_{-0.2} \times 10^{-13}$
        & $2.7^{+0.3}_{-1.3} \times 10^{-13}$\\[5pt]
$f^{\rm un}_{0.3-10}$ & \ \ \ \ $4.3^{+1.0}_{-0.7} \times 10^{-13}$ 
        & $4.4^{+1.1}_{-0.8} \times 10^{-13}$
        & $4.1^{+1.0}_{-0.6} \times 10^{-13}$
        & $4.3^{+1.5}_{-0.8} \times 10^{-13}$\\[5pt]
\hline\\[-5pt]
$\chi^2_{\nu}$ &  \ \ \ \  $0.99$ (174.1/175) & $1.00$ (174.7/175)
      & $1.00$ (174.8/175) & $1.00$ (175.3/175)  
\end{tabular}
\end{center}
\caption{Best-fitting spectral parameters for the 2004 {\it XMM-Newton}/EPIC 
observation. In addition to the intrinsic column density listed here, 
we included a line-of-sight column density of $2 \times 10^{20}$ cm$^{-2}$.
Errors are 90\% confidence level for 1 interesting parameter. The fitting parameter 
for the {\it zamp} model is $\left(L/L_{\rm Edd}\right)$, but we have also 
listed the corresponding value of $kT_{\rm eff}$ for an easier comparison 
with the other models.}
\label{tab:a}
\end{table*}

\begin{table*}
\begin{center}
\begin{tabular}{lrrrr}\hline
\multicolumn{5}{c}{{\bf Second XMM observation}} \\
\hline
 & wabs*(bb+po) & wabs*(diskbb+po) & wabs*(zamp+po) & wabs*(nsa+po)  \\
\hline
Parameter & \multicolumn{4}{c}{Value}\\
\hline\\[-5pt]
$N_{H,\rm int}$ & \ \ \ \ $<0.5 \times 10^{20}$ cm$^{-2}$ & \ \ \ \ 
        $<1.3 \times 10^{20}$ cm$^{-2}$ & \ \ \ \ 
        $<0.9 \times 10^{20}$ cm$^{-2}$ & \ \ \ \ 
        $0.9^{+1.3}_{-0.9} \times 10^{20}$ cm$^{-2}$ 
        \\[5pt]
$kT_{\rm bb}$  
        & $0.133^{+0.003}_{-0.003}$ keV &&&
        \\[5pt]
$N_{\rm bb}$  
        & $3.33^{+0.22}_{-0.35}  \times 10^{-6}$ &&&\\[5pt]
$kT_{\rm dbb}$ & 
        & $0.186^{+0.005}_{-0.004}$ keV &&
        \\[5pt]
$N_{\rm dbb}$ & 
        & $20.4^{+13.9}_{-2.0}$ &&\\[5pt]
$\log \left(L/L_{\rm Edd}\right)$ &&&  $-6.02^{+0.05}_{-0.01}$ &\\[5pt]
$N_{\rm zam}$ &&&  $4.67^{+0.24}_{-0.29} \times 10^{-5}$ & \\[5pt]
$kT_{\rm eff}$ & & & [$0.060^{+0.002}_{-0.001}$ keV]
        & $0.055^{+0.002}_{-0.002}$ keV
        \\[5pt]
$N_{\rm nsa}$ & & &
        & $5.5^{+0.4}_{-1.3} \times 10^{-7}$ 
        \\[5pt]
$\Gamma$ & $2.62^{+0.13}_{-0.18}$
         & $2.03^{+0.23}_{-0.22}$
         & $1.89^{+0.17}_{-0.22}$
         & $1.82^{+0.18}_{-0.16}$
         \\[5pt]
$N_{\rm po}$ & $3.4^{+0.4}_{-0.4} \times 10^{-5}$
         & $1.9^{+0.5}_{-0.4} \times 10^{-5}$
         & $1.5^{+0.3}_{-0.3} \times 10^{-5}$
         & $1.4^{+0.4}_{-0.3} \times 10^{-5}$
         \\[5pt]
\hline\\[-5pt]
$f_{0.3-10}$ & \ \ \ \ $3.2^{+0.1}_{-0.1} \times 10^{-13}$ 
        & $3.4^{+0.1}_{-0.1} \times 10^{-13}$
        & $3.4^{+0.1}_{-0.1} \times 10^{-13}$
        & $3.4^{+0.3}_{-0.6} \times 10^{-13}$\\[5pt]
$f^{\rm un}_{0.3-10}$ & \ \ \ \ $3.8^{+0.2}_{-0.1} \times 10^{-13}$ 
        & $4.0^{+0.2}_{-0.1} \times 10^{-13}$
        & $3.9^{+0.3}_{-0.1} \times 10^{-13}$
        & $4.2^{+0.4}_{-0.2} \times 10^{-13}$\\[5pt]
\hline\\[-5pt]
$\chi^2_{\nu}$ &  \ \ \ \  $0.94$ (356.5/381) & $0.93$ (352.6/381)
      & $0.92$ (351.7/381) & $0.93$ (356.2/381)  
\end{tabular}
\end{center}
\caption{As in Table 1, for the 2008 {\it XMM-Newton}/EPIC 
observation.} 
\label{tab:b}
\end{table*}

\subsection{{\it XMM-Newton} results}

For both XMM1 and XMM2, we fitted pn and MOS spectra simultaneously, 
leaving a free normalization constant bewteeen the three instruments.
{\it A priori}, there might have been significant discrepancies due 
to the pn chip gap cutting across the source (in XMM1), or to the fact 
that pn and MOS were in different modes (in XMM2). In fact, 
the pn and MOS events processed and extracted with SAS Version 9.0.0 
turned out to be consistent both in spectral shape 
and normalization (within 3\%), for both observations.
This is not the case when pn and MOS event files are processed 
with earlier versions of the SAS (for example, version 6.6.0 was used 
for the pipeline-processed files in the public archives);  
in particular, for those older versions there is an $\approx 10$\% discrepancy between 
pn and MOS below 0.5 keV, which makes it difficult to get strong constraints 
on the soft thermal component. This may be the reason why 
\citet[][their Fig.~2]{far09} chose to ignore all pn data 
below 0.5 keV in XMM2, and MOS2 data below 0.4 keV in XMM1, 
without providing any explanations.

The first result of our analysis is that the XMM1 spectrum 
is dominated by a power-law component, while the XMM2 spectrum 
is dominated by the thermal component: this is in general agreement 
with the results of \citet{far09} and \citet{god09}, 
and confirms their finding that there was a significant spectral 
transition between XMM1 and XMM2.
However, unlike \citet{far09}, we find that the XMM1 spectral fit  
is improved by the addition of a soft thermal component
(F-test significance $\approx 95$\% for each of the four 
thermal models used in this work). In XMM1, the thermal component carries 
$\approx 1/3$ of the emitted luminosity in the $0.3$--$10$ keV band; 
in XMM2, it accounts for almost $80$\% of the emitted X-ray luminosity.
It is possible to introduce additional parameters, for example using 
low-metallicity or ionized absorbers, to obtain statistically 
equivalent fits without a soft thermal component for XMM1.
However, the price to pay is that a steeper power-law slope 
is required (photon index $\Gamma > 3$, unusual for an accreting 
compact object). Besides, a combination of thermal and power-law 
components is seen in all other observations of this source 
(XMM2 and {\it Swift}); so, there are no compelling reasons 
for forcing the first spectrum to be fitted by a simple power law.
 
When both XMM1 and XMM2 are fitted with a thermal plus power-law model, 
we find (Tables 2,3) that the un-absorbed flux is the same 
for both observations, 
$f^{\rm un}_{0.3-10} \approx 4 \times 10^{-13}$ erg cm$^{-2}$ s$^{-1}$ 
(corresponding to a luminosity $\approx 4 \times 10^{41}$ erg s$^{-1}$ 
at a distance of 91 Mpc), in contrast with the findings of \citet{far09}. 
This is important because 
it means that the spectral transition between XMM1 and XMM2 
was more likely due to a change in the relative fraction 
of thermal/non-thermal photon output at constant luminosity, 
rather than to a change in the accretion rate. 
There are subsequent {\it Swift} observations when 
the un-absorbed flux does increase, reaching $\approx 8 \times 10^{-13}$ erg 
cm$^{-2}$ s$^{-1}$ (Table 4), but their spectral properties 
look very different from those seen in XMM1 (Section 3.3).


The colour temperature of the thermal component is $\approx 0.13$ keV 
for both XMM1 and XMM2, when fitted with a simple blackbody model, 
while the effective temperature from the {\it zamp} and {\it nsa} models 
is $\approx 0.06$ keV.  This is expected, because of the hardening effect 
on the photon spectrum emerging from the NS atmosphere. It implies  
a hardening factor $\approx 2.7$, which is similar to what was calculated 
by \citet[][their Table 1]{zam95} for this range of luminosities 
($L \approx 10^{-6} L_{\rm Edd} \approx 2 \times 10^{32}$ erg s$^{-1}$, 
in the NS scenario). 

The best-fitting intrinsic column density is higher in XMM1 
($\approx 4 \times 10^{20}$ cm$^{-2}$) than in XMM2 
($< 10^{20}$ cm$^{-2}$), which is at least qualitatively 
consistent with the results of \citet{far09}. We suggest that this 
may not reflect a true physical change in the source. Our X-ray 
spectral models of XMM1 are dominated by the phenomenological 
power-law component, which requires intrinsic absorption 
to avoid divergence at soft energies. We have tried replacing 
the power-law model with more complex Comptonization models, 
where the power-law-like component has an intrinsic turnover 
at low energies, around the temperature of the seed photons.
We obtain equally acceptable fits (for example, $\chi^2_{\nu} = 172.0/175$ 
for a {\it comptt} model) with similar seed-photon temperatures 
$\approx 0.1$ keV and no intrinsic absorption.
On the other hand, XMM2 and the {\it Swift} spectra (Section 3.3) 
are dominated by the thermal component at low energies, 
and do not require artificial addition of intrinsic absorption.

The next step is to determine whether the thermal emission  
is better fitted by a single-temperature or multi-temperature
component. We find that all four models give equivalent fits
for XMM1 and XMM2. Taking into account all five sets of spectral fits 
(both {\it XMM-Newton} observations, and the three grouped 
datasets from {\it Swift}), there is a hint that a single-temperature 
blackbody may give a slightly worse fit than broader thermal components 
(Tables 2--6), but longer observations will be necessary 
to test this suggestion.
In any case, there is no statistical difference between 
disk-blackbody models (most suitable to an intermediate-mass 
BH scenario) and NS atmosphere models. X-ray spectroscopy alone 
cannot rule out either scenario.

\subsection{{\it Swift} results}

Individual {\it Swift}/XRT observations do not have enough counts 
to allow two-component spectral fits and to provide any 
constraints on the relative contribution and temperature 
of the thermal component. To get around this problem, 
we examined the {\it Swift}/XRT lightcurve and grouped the observations 
into four bands, at very high (``A''), high (``B''), intermediate (``C'') 
and low (``D'') count rates (Figure 3). We then coadded the spectra from all 
the observations in each band. Band D is still too faint 
for two-component spectral fitting, and we will not discuss it here.
For bands A, B and C we used the same four spectral models 
applied to XMM1 and XMM2 (Tables 4--6). 
The total exposure time of the coadded band-A spectrum is 19.0 ks;
for band B, 22.3 ks; for band C, 67.3 ks.

We found that a thermal component is required for the combined spectrum 
of every band. In fact, the band-A spectrum (corresponding to the outburst peak 
in 2009 August) is consistent with only a thermal component, without a power-law, 
although the upper limit to the power-law normalization is not very constraining. 
The fractional power-law contribution becomes more important for the band-B and band-C 
spectra at lower luminosities, mainly because the thermal component declines. 
Spectral parameters and un-absorbed flux of the band-C spectrum are very similar 
to those of XMM2, although at lower signal to noise.

\begin{table*}
\begin{center}
\begin{tabular}{lrrrr}\hline
\multicolumn{5}{c}{{\bf Band-A Swift observations}} \\
\hline
 & wabs*(bb+po) & wabs*(diskbb+po) & wabs*(zamp+po) & wabs*(nsa+po)  \\
\hline
Parameter & \multicolumn{4}{c}{Value}\\
\hline\\[-5pt]
$N_{H,\rm int}$ & $0$ (fixed) & \ \ \ \ 
        $0$ (fixed) & \ \ \ \ 
        $0$ (fixed) & \ \ \ \ 
        $0$ (fixed) 
        \\[5pt]
$kT_{\rm bb}$  
        & $0.18^{+0.03}_{-0.02}$ keV &&&
        \\[5pt]
$N_{\rm bb}$  
        & $9.1^{+2.7}_{-2.3}  \times 10^{-6}$ &&&\\[5pt]
$kT_{\rm dbb}$ & 
        & $0.283^{+0.018}_{-0.021}$ keV &&
        \\[5pt]
$N_{\rm dbb}$ & 
        & $8.9^{+2.8}_{-2.5}$ &&\\[5pt]
$\log \left(L/L_{\rm Edd}\right)$ &&&  $-5.07^{+0.13}_{-0.13}$ &\\[5pt]
$N_{\rm zam}$ &&&  $25.2^{+2.4}_{-2.4} \times 10^{-5}$ & \\[5pt]
$kT_{\rm eff}$ & & & [$0.104^{+0.008}_{-0.008}$ keV]
        & $0.094^{+0.009}_{-0.009}$ keV
        \\[5pt]
$N_{\rm nsa}$ & & &
        & $1.2^{+0.6}_{-0.4} \times 10^{-7}$ 
        \\[5pt]
$\Gamma$ & $2.2^{+1.2}_{-0.9}$
         & $2.0$ (fixed)
         & $2.0$ (fixed)
         & $2.0$ (fixed)
         \\[5pt]
$N_{\rm po}$ & $5.9^{+4.8}_{-5.4} \times 10^{-5}$
         & $< 4.6 \times 10^{-5}$
         & $< 5.1 \times 10^{-5}$
         & $< 4.4 \times 10^{-5}$
         \\[5pt]
\hline\\[-5pt]
$f_{0.3-10}$ & \ \ \ \ $8.6^{+2.9}_{-0.9} \times 10^{-13}$ 
        & $7.8^{+0.2}_{-0.6} \times 10^{-13}$
        & $7.7^{+0.6}_{-0.7} \times 10^{-13}$
        & $7.7^{+1.3}_{-3.2} \times 10^{-13}$\\[5pt]
$f^{\rm un}_{0.3-10}$ & \ \ \ \ $9.7^{+2.4}_{-0.9} \times 10^{-13}$ 
        & $8.9^{+1.1}_{-0.2} \times 10^{-13}$
        & $8.7^{+1.3}_{-0.2} \times 10^{-13}$
        & $8.8^{+1.2}_{-0.2} \times 10^{-13}$\\[5pt]
\hline\\[-5pt]
$\chi^2_{\nu}$ &  \ \ \ \  $0.76$ (14.5/19) & $0.74$ (15.6/21)
      & $0.67$ (14.2/21) & $0.69$ (14.6/21)  
\end{tabular}
\end{center}
\caption{Best-fitting spectral parameters for the coadded {\it Swift}/XRT 
observations at the peak of the 2009 August outburst. See Figure 3 
for our definition of band A. 
We fixed the intrinsic column density to zero (it converges to zero even 
when left as a free fitting parameter). In addition, we included a line-of-sight 
column density of $2 \times 10^{20}$ cm$^{-2}$.
Errors are 90\% confidence level for 1 interesting parameter.}
\label{tab:c}
\end{table*}

\begin{table*}
\begin{center}
\begin{tabular}{lrrrr}\hline
\multicolumn{5}{c}{{\bf Band-B Swift observations}} \\
\hline
 & wabs*(bb+po) & wabs*(diskbb+po) & wabs*(zamp+po) & wabs*(nsa+po)  \\
\hline
Parameter & \multicolumn{4}{c}{Value}\\
\hline\\[-5pt]
$N_{H,\rm int}$ & $0$ (fixed) & \ \ \ \ 
        $0$ (fixed) & \ \ \ \ 
        $0$ (fixed) & \ \ \ \ 
        $0$ (fixed) 
        \\[5pt]
$kT_{\rm bb}$  
        & $0.14^{+0.02}_{-0.02}$ keV &&&
        \\[5pt]
$N_{\rm bb}$  
        & $4.9^{+2.3}_{-2.4}  \times 10^{-6}$ &&&\\[5pt]
$kT_{\rm dbb}$ & 
        & $0.194^{+0.027}_{-0.024}$ keV &&
        \\[5pt]
$N_{\rm dbb}$ & 
        & $26.5^{+20.2}_{-12.2}$ &&\\[5pt]
$\log \left(L/L_{\rm Edd}\right)$ &&&  $-5.92^{+0.27}_{-0.17}$ &\\[5pt]
$N_{\rm zam}$ &&&  $8.0^{+3.2}_{-3.7} \times 10^{-5}$ & \\[5pt]
$kT_{\rm eff}$ & & & [$0.064^{+0.10}_{-0.06}$ keV]
        & $0.060^{+0.009}_{-0.007}$ keV
        \\[5pt]
$N_{\rm nsa}$ & & &
        & $5.3^{+5.5}_{-2.6} \times 10^{-7}$ 
        \\[5pt]
$\Gamma$ & $2.8^{+0.5}_{-0.4}$
         & $2.2^{+1.0}_{-1.6}$
         & $2.0$ (fixed)
         & $2.0$ (fixed)
         \\[5pt]
$N_{\rm po}$ & $5.4^{+2.8}_{-2.5} \times 10^{-5}$
         & $2.7^{+4.2}_{-2.6} \times 10^{-5}$
         & $2.1^{+1.0}_{-1.2} \times 10^{-5}$
         & $2.0^{+1.1}_{-1.1} \times 10^{-5}$
         \\[5pt]
\hline\\[-5pt]
$f_{0.3-10}$ & \ \ \ \ $5.0^{+1.0}_{-0.7} \times 10^{-13}$ 
        & $5.1^{+2.0}_{-0.7} \times 10^{-13}$
        & $5.1^{+0.8}_{-0.5} \times 10^{-13}$
        & $5.1^{+1.0}_{-3.0} \times 10^{-13}$\\[5pt]
$f^{\rm un}_{0.3-10}$ & \ \ \ \ $5.9^{+0.4}_{-0.2} \times 10^{-13}$ 
        & $6.0^{+0.7}_{-0.2} \times 10^{-13}$
        & $6.0^{+0.3}_{-0.3} \times 10^{-13}$
        & $6.1^{+0.3}_{-0.4} \times 10^{-13}$\\[5pt]
\hline\\[-5pt]
$\chi^2_{\nu}$ &  \ \ \ \  $1.04$ (21.8/21) & $1.03$ (21.7/21)
      & $0.99$ (21.8/22) & $1.00$ (22.0/22)  
\end{tabular}
\end{center}
\caption{As in Table 4, for the {\it Swift}/XRT 
observations during the decline from the 2009 August outburst. 
(as defined in Figure 3).}
\label{tab:d}
\end{table*}

\begin{table*}
\begin{center}
\begin{tabular}{lrrrr}\hline
\multicolumn{5}{c}{{\bf Band-C Swift observations}} \\
\hline
 & wabs*(bb+po) & wabs*(diskbb+po) & wabs*(zamp+po) & wabs*(nsa+po)  \\
\hline
Parameter & \multicolumn{4}{c}{Value}\\
\hline\\[-5pt]
$N_{H,\rm int}$ & $0$ (fixed) & \ \ \ \ 
        $0$ (fixed) & \ \ \ \ 
        $0$ (fixed) & \ \ \ \ 
        $0$ (fixed) 
        \\[5pt]
$kT_{\rm bb}$  
        & $0.14^{+0.02}_{-0.01}$ keV &&&
        \\[5pt]
$N_{\rm bb}$  
        & $3.0^{+1.0}_{-1.3}  \times 10^{-6}$ &&&\\[5pt]
$kT_{\rm dbb}$ & 
        & $0.204^{+0.022}_{-0.022}$ keV &&
        \\[5pt]
$N_{\rm dbb}$ & 
        & $12.5^{+7.2}_{-4.3}$ &&\\[5pt]
$\log \left(L/L_{\rm Edd}\right)$ &&&  $-5.78^{+0.21}_{-0.26}$ &\\[5pt]
$N_{\rm zam}$ &&&  $5.2^{+1.3}_{-1.8} \times 10^{-5}$ & \\[5pt]
$kT_{\rm eff}$ & & & $[0.069^{+0.009}_{-0.010}$ keV]  
        & $0.066^{+0.008}_{-0.007}$ keV
        \\[5pt]
$N_{\rm nsa}$ & & &
        & $2.1^{+2.0}_{-0.8} \times 10^{-7}$ 
        \\[5pt]
$\Gamma$ & $2.2^{+1.2}_{-0.9}$
         & $1.5^{+0.8}_{-1.0}$
         & $1.4^{+0.8}_{-0.9}$
         & $1.2^{+1.1}_{-0.9}$
         \\[5pt]
$N_{\rm po}$ & $3.4^{+1.8}_{-1.3} \times 10^{-5}$
         & $1.3^{+2.0}_{-1.0} \times 10^{-5}$
         & $1.1^{+1.9}_{-0.8} \times 10^{-5}$
         & $0.9^{+1.8}_{-0.6} \times 10^{-5}$
         \\[5pt]
\hline\\[-5pt]
$f_{0.3-10}$ & \ \ \ \ $3.2^{+0.3}_{-0.4} \times 10^{-13}$ 
        & $3.4^{+0.3}_{-0.9} \times 10^{-13}$
        & $3.4^{+0.1}_{-0.6} \times 10^{-13}$
        & $3.5^{+0.5}_{-2.0} \times 10^{-13}$\\[5pt]
$f^{\rm un}_{0.3-10}$ & \ \ \ \ $3.6^{+0.2}_{-0.2} \times 10^{-13}$ 
        & $3.9^{+0.3}_{-0.2} \times 10^{-13}$
        & $3.9^{+0.3}_{-0.3} \times 10^{-13}$
        & $4.0^{+0.3}_{-0.3} \times 10^{-13}$\\[5pt]
\hline\\[-5pt]
$\chi^2_{\nu}$ &  \ \ \ \  $0.77$ (25.6/33) & $0.68$ (22.5/33)
      & $0.68$ (22.6/33) & $0.68$ (22.4/33)  
\end{tabular}
\end{center}
\caption{As in Table 4, for the {\it Swift}/XRT 
observations of luminosity band C (as defined in Figure 3).}
\label{tab:e}
\end{table*}

Putting together the {\it Swift} and {\it XMM-Newton} spectral results, 
we find two possible trends, which will have to be tested by longer 
observations. First, the relative contribution of the power-law component 
seems to decrease at higher luminosities (Figure 5). This can be explained 
in the framework of BH accretion, if the Comptonizing region (responsible 
for the power-law component) collapses to an optically thick disk, 
at high mass accretion rates. In the framework of weakly accreting NSs, 
our {\it zamp} and {\it nsa} spectral fits suggest an un-absorbed $0.3$--$10$ keV 
luminosity varying between $\approx 2$--$5 \times 10^{32}$ erg s$^{-1}$ (Table 7). 
In this luminosity range, accreting NSs are known to have a power-law 
and a thermal X-ray component, with the relative power-law contribution decreasing 
as the luminosity increases \citep[][their Fig.~5]{jon04}. 
Thus, the X-ray spectral evolution of HLX1 also has some apparent 
similarities with the behaviour of weakly accreting NSs. 

The second possible trend is an increase in the temperature of the thermal component 
at higher luminosities (Figure 6). This is consistent with optically-thick emission 
from a surface of approximately fixed size, which gets hotter perhaps as a result 
of enhanced accretion rate; it rules out the alternative possibility of 
a change in the soft X-ray luminosity due to an expanding photosphere.
But this scenario is equally applicable to emission from the disk around 
a BH, or from a NS surface: in both cases, we expect $L \sim T_{\rm eff}^4$ 
when the size of the X-ray emitting region is fixed.

\begin{figure}
\begin{center}
\psfig{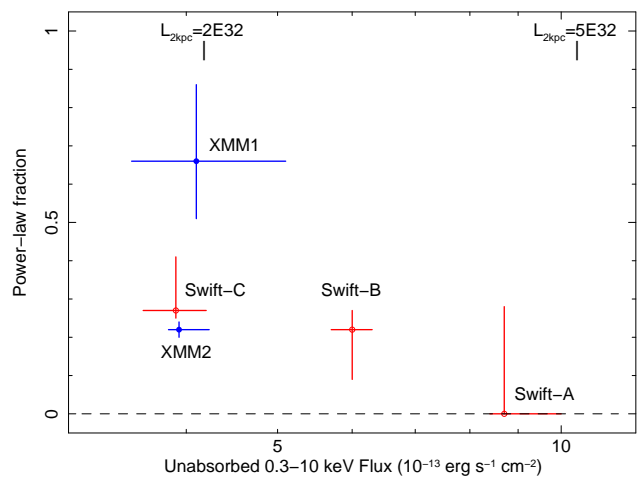}
\end{center}
\caption{Fractional contribution of the power-law component 
to the un-absorbed $0.3$--$10$ keV flux. The values and error bars 
have been calculated with the {\it zamp} model, but very similar
results are obtained with the other three models. The luminosity 
markers are for a distance of 2 kpc (NS scenario).}
\label{f5}
\end{figure}

\begin{figure}
\begin{center}
\psfig{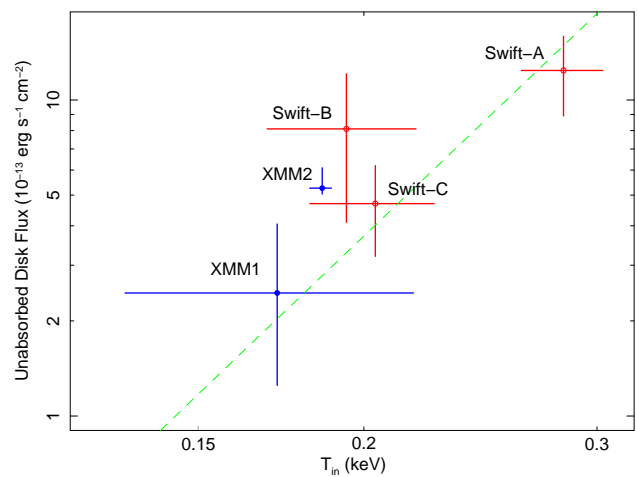}
\end{center}
\caption{Relation between peak colour temperature of the best-fitting 
{\it diskbb} model, and total un-absorbed flux in the {\it diskbb} component.
The dashed line is not a fit to the data: it marks the location of 
the $L \propto T^4$ correlation, expected for optically-thick 
thermal emission from a region of constant area at increasing 
accretion rate (inner disk or neutron star surface).
Very similar relations between thermal luminosity and temperature 
are obtained with the other three spectral models.}
\label{f6}
\end{figure}

\section{Short-term X-ray variability}

The combined EPIC lightcurve for XMM2 does not appear particularly 
remarkable (Figure 7), but it clearly suggests some 
short-term variability. To quantify such variability, 
we used the background-subtracted pn lightcurve binned 
to a time resolution of 1 s. We obtain a 
Kolmogorov-Smirnoff probability of constancy 
$\approx 8 \times 10^{-3}$, and $(53 \pm 5)\%$ rms fractional variation 
in excess of the Poisson level.
We applied the same analysis to a combined EPIC lightcurve binned 
to 10-s intervals, obtaining a Kolmogorov-Smirnoff probability 
of constancy $1.3 \times 10^{-2}$ and $(34 \pm 2)\%$ rms fractional variation.

\begin{figure}
\vspace{-0.35cm}
\begin{center}
\psfig{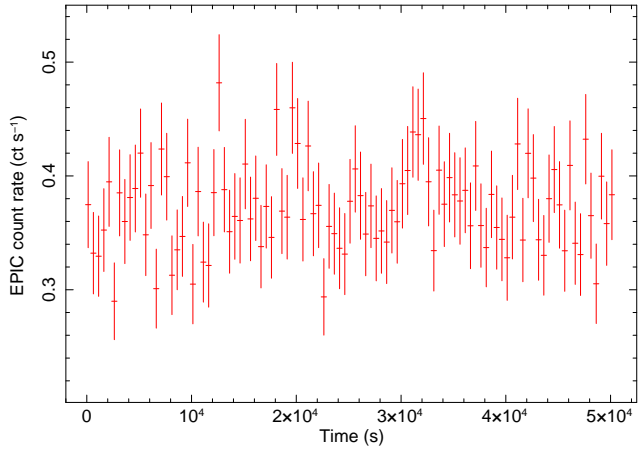}
\end{center}
\caption{Combined pn + MOS lightcurve in the $0.2$--$12$ keV 
range, binned to a time resolution of 500 s. }
\label{f7}
\end{figure}

We searched for characteristic periods in the pn and combined EPIC 
lightcurves, by folding the data to a range of periods 
({\it efsearch} in {\footnotesize{FTOOLS}}). 
We do not find any strong or well defined period; however, 
there is a weak, quasi-periodic modulations 
with characteristic periods $\approx 5300$--$5600$ s (Figure 8).
The power associated with this modulation or range of characteristic 
modulations is small, barely about the noise level, as can be seen 
from the power spectral density plot (Figure 9). At higher frequencies, 
the power spectral density is consistent with white noise.

\begin{figure}
\begin{center}
\psfig{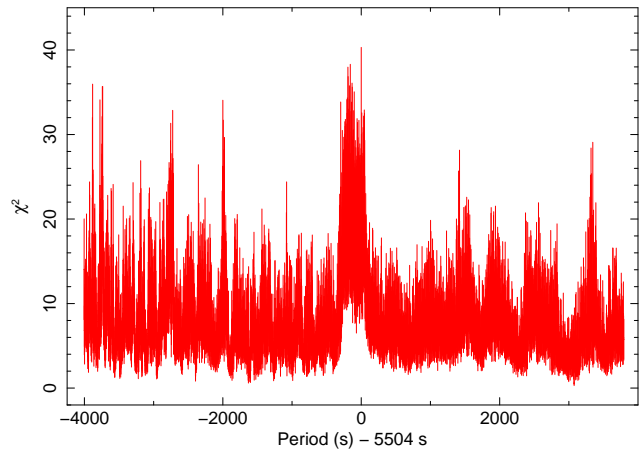}
\end{center}
\caption{A search for periodicities in the combined EPIC lightcurve 
suggests the presence of weak, quasi-periodic modulations 
with characteristic periods $\approx 5300$--$5600$ s, and possibly 
other weak features in the $1000$--$10,000$ s range.}
\label{f8}
\end{figure}

\begin{figure}
\begin{center}
\psfig{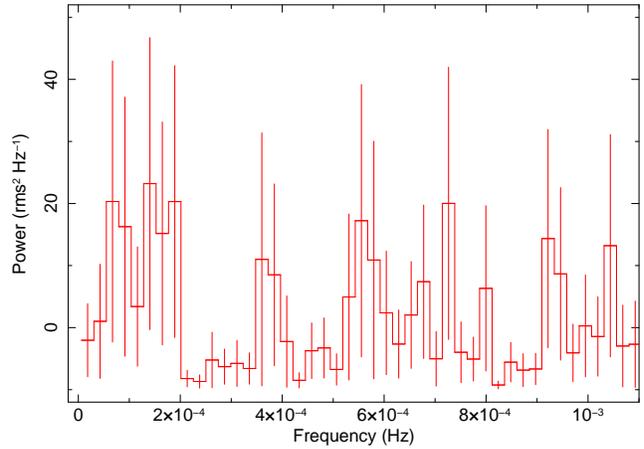}
\end{center}
\caption{Low-frequency section of the power spectral density  
for the combined EPIC lightcurve. The fitted white noise level 
has been subtracted (normalization parameter = $-2$ in 
the FTOOLS task {\it powspec}).}
\label{f9}
\end{figure}

\section{Discussion and Conclusions}

We have re-examined the X-ray spectral evolution of HLX1, 
using two {\it XMM-Newton} observations from 2004 and 2008, 
and a series of {\it Swift} observations over 2008--2010.
In general, a two-component model consisting of power law plus 
soft thermal component provides a good fit to all the spectra.
The first XMM observation is dominated by the power-law emission 
(contributing $\approx 2/3$ of the flux) but a thermal component 
with a colour temperature $\approx 0.13$ keV is also significantly 
detected. Conversely, at the peak of the {\it Swift} outburst, 
the X-ray spectrum is dominated by a thermal component 
with a colour temperature $\approx 0.2$ keV.

Although the relative contributions of thermal component and power-law 
are significantly different between the {\it XMM-Newton} observations, 
the un-absorbed flux is similar, and a factor of two lower than 
previously claimed \citep{far09}. We attribute this discrepancy to our 
processing of the EPIC event files with the latest version 
of the SAS, which may provide more accurate results at energies $\la 0.5$ keV. 
With our re-analysis, pn and MOS now give identical spectral 
parameters and normalizations consistent within 3\%, even below 0.5 keV.
As an aside, we note that the spectral difference between 
the two {\it XMM-Newton} observations was visually exaggerated  
in \citet{far09} by their choice of plotting two unfolded spectra based 
on different models (simple power-law, and disk-blackbody plus power-law).

For the thermal component, there is no statistical difference between 
disk-blackbody models (most suitable to an intermediate-mass 
BH scenario) and NS atmosphere models. X-ray spectroscopy alone 
cannot rule out either scenario.
The {\it diskbb} model is normalized in terms of apparent inner-disk 
radius (assumed to coincide with the radius of the innermost stable circular 
orbit, at high accretion rates). Normalization values $\sim 10$--$20$ (Tables 2, 3) 
agree with the findings of \citet{far09}, and would correspond to 
a BH mass $\sim 10^4 M_{\odot}$ at the distance of 91 Mpc, 
even higher than conservatively estimated from the Eddington limit. 
The small differences (less than a factor of 2) between the fitted radii over 
the five sets of observations may be due to the fact that the inner disk 
is not always extending precisely to the innermost stable circular orbit, 
as well as to other effects such as the degree of Comptonization 
and spectral hardening. If HLX1 belongs to ESO243$-$49, our spectral fits imply 
emitted luminosities $\approx 4 \times 10^{41}$ erg s$^{-1}$ 
in the $0.3$--$10$ keV band, during 
the 2004 and 2008 {\it XMM-Newton} observations, rising to 
$\approx 10^{42}$ erg s$^{-1}$ at the peak of the August 2009 outburst.

The {\it zamp} and {\it nsa} models are normalized in terms of distance 
to the source, assuming uniform emission from the surface of a $1.4 M_{\odot}$ NS with 
a radius of $12.4$ km. Both models suggest a distance $\approx 1.5$--$3$ kpc 
(Table 7). The small discrepancies between the fitted distances over 
the five sets of observations may be due to the fact that the emission 
is never perfectly isotropic from the whole NS surface.
If HLX1 is a weakly accreting NS, its $0.3$--$10$ keV luminosity varies 
between $\approx (2$--$5) \times 10^{32}$ erg s$^{-1}$, 
that is $\sim 10^{-6} L_{\rm Edd}$. This requires accretion rates 
$\dot{M} \sim 10^{-13} M_{\odot}$ yr$^{-1}$.
At a distance of 2 kpc, the optical 
counterpart found by \citet{sor10} would be a main-sequence M star.
We also note that any direct X-ray emission from the M dwarf donor 
would be several orders of magnitude below our detection limit: 
the maximum X-ray luminosity 
of such stars is $\approx 10^{29}$ erg s$^{-1}$ \citep{jam00,sch95}.

The increase in the temperature and luminosity of the thermal component 
during the 2009 outburst, and the decrease of the fractional contribution 
of the power-law component, may be consistent with both a disk-dominated, high-state BH 
and with a quiescent low-mass NS X-ray binary (in particular, it is typically seen 
in NS X-ray binaries 
with X-ray luminosities $10^{32} \la L_{\rm X} \la 10^{33}$ erg s$^{-1}$).
In HLX1, the power-law component itself is not constant, as we can see  
by comparing XMM1 with XMM2 and with the very faint Band-D {\it Swift} observations.
In weakly accreting NSs, it was suggested, based on the sample of \citet{jon04}, 
that the power-law component could represent a constant baseline luminosity 
at $\approx 10^{32}$ erg s$^{-1}$ for most sources, while the thermal 
component increases with flux. But other quiescent 
NS X-ray binaries contradict this picture, because they decline to  
luminosities (including the power-law component) $\la 5 \times 10^{31}$ 
erg s$^{-1}$ \citep{hei05} or even 
$\la 3 \times 10^{30}$ erg s$^{-1}$ \citep[1H 1905$+$000:][]{jon07}.
Evidence of flux variability in both the thermal and power-law components 
is sometimes clearly seen in quiescent NSs \citep{cac05,jon05}; but it is still not clear how 
the variability of the two components may be related, because we do not know 
the physical origin of the faint power-law component.

\begin{table}
\begin{center}
\begin{tabular}{lccc}
\hline
Observation & $f^{\rm un}_{0.3-10}$  & $d_{\rm zamp}$ & $d_{\rm nsa}$ \\
 & ($10^{-13}$ erg cm$^{-2}$ s$^{-1}$) & (kpc) & (kpc) \\
\hline\\[-5pt]
XMM1 & $4.1^{+1.0}_{-0.6}$ & $2.4^{+3.0}_{-1.3}$ & $2.4^{+2.6}_{-1.9}$ \\[5pt]
XMM2 & $3.9^{+0.3}_{-0.1}$ & $1.8^{+0.1}_{-0.1}$ & $1.3^{+0.2}_{-0.1}$ \\[5pt]
Swift-A & $8.7^{+1.3}_{-0.2}$ & $3.6^{+0.6}_{-0.9}$ & $2.9^{+0.6}_{-0.5}$ \\[5pt]
Swift-B & $6.0^{+0.3}_{-0.3}$ & $1.6^{+0.5}_{-0.5}$ & $1.4^{+0.5}_{-0.4}$ \\[5pt]
Swift-C & $3.9^{+0.3}_{-0.3}$ & $2.5^{+0.8}_{-0.7}$ & $2.2^{+0.6}_{-0.6}$ \\[5pt]
\hline
\end{tabular}
\end{center}
\caption{Best-fitting un-absorbed fluxes and distance to HLX1 in the NS scenario, 
from the normalization parameters of the {\it zamp} and {\it nsa} models (Tables 2--6). }
\end{table}

The X-ray outburst and spectral variability seen in 2009 
may be explained as ``canonical'' state transitions of BH accretion.
But they can also be interpreted in terms of sporadic accretion 
onto the surface of a NS with a low-mass donor.
Material may accumulate at the magnetospheric radius, 
then be sporadically "flushed" down towards the NS, 
either because of changes in the mass transfer rate 
from the donor star, or changes in the alignment of 
non-dipole components of the NS field.

The presence of a variable power-law component (dominant contribution 
in XMM1, less important in XMM2 and even less in the bright {\it Swift} state), 
with a variable photon index $\approx 2$--$3$, does 
not unambiguosly identify BH or NS accretion.
The power-law emission may come from inverse-Compton scattering 
of the soft X-ray photons by more energetic electrons in a hot corona, 
located either above the inner accretion disk or the NS surface.
Observationally, the NS X-ray binary Aql X-1 shows X-ray spectral 
variability in quiescence, probably due to variable residual accretion: 
it has a soft thermal component with effective temperature 
varying between $\approx 0.11$--$0.13$ keV, and a power-law component 
with photon index varying between $\approx 1.5$ and $\approx 4$, 
or disappearing altogether \citep{rut02,cam03}.
Cen X-4 is another quiescent NS X-ray binaries well fitted by a soft thermal 
component ($kT \approx 0.16$ keV) plus power-law ($\Gamma \sim 2$) 
when its X-ray luminosity was $\sim$ a few $10^{32}$ erg s$^{-1}$ 
\citep{cam04,cam98,asa96}. Other quiescent NS X-ray binaries 
with comparable contributions from a soft thermal and a power-law component 
are listed in \citet{jon04}. Quiescent low-mass X-ray binary 
with X-ray luminosities $\sim 10^{32}$ erg s$^{-1}$, a thermal component 
and a steep ($\Gamma > 2$) power-law component were observed 
in the globular cluster 47 Tuc \citep[W37 and X4:][]{hei05}.
Short-term X-ray variability was also seen in some observations 
of Aql X-1 \citep[at 32\% rms;][]{rut02} and Cen X-4 \citep[at 45\% rms;][]{cam04}. 
In fact, it was suggested \citep{hei03a,hei03b} that the strength of the power-law 
component and the presence of intrinsic short-term variability 
in quiescent NS low-mass X-ray binaries are two indicators 
of continued low-level accretion. The (speculative) detection 
of weak modulations around $\approx 5300$--$5600$ s 
can have many explanations and does not uniquely 
identify an intermediate-mass BH. For example, it is also 
the orbital period of an M4 main-sequence star filling its Roche lobe, 
from the well-known period-density relation in binary systems \citep{fra02}.


In conclusion, our X-ray spectral and timing analysis has provided more accurate constraints 
on the un-absorbed flux, degree of variability, relative thermal/non-thermal 
contribution, and temperature of the thermal component 
from HLX1. However, X-ray properties alone are not sufficient 
to rule out either of the competing models previously suggested for this source 
(old NS with a low-mass donor in the Galactic halo, or intermediate-mass BH in ESO\,243$-$49).
This is because the thermal component is equally consistent 
with emission from an accretion disk around a BH or from the NS surface, 
and the presence of an additional power-law component does not unambiguously identify 
the BH scenario, either. The X-ray spectral properties are consistent with 
an intermediate-mass BH in the high or very high state, but also with 
a quiescent NS with low-level (and variable) residual accretion. 
X-ray variability properties are also consistent with both scenarios.
We suggest that, regardless of the true nature of HLX1, its X-ray properties  
do not yet provide a unique observational signature for the identification of 
the new class of intermediate-mass BHs. Therefore, the identification of HLX1 
as an intermediate-mass BH must rely on its properties in other bands, 
for example from optical spectroscopy \citep{far10}. 


\section*{Acknowledgments}

This work made use of data supplied by the UK Swift Science Data Centre 
at the University of Leicester. We thank Sergio Campana for providing
the {\footnotesize{XSPEC}} implementation of our models.
We thank Rosanne Di Stefano, Jeanette Gladstone, Alister W. Graham, George Hau, 
Craig Heinke, Albert Kong, Jifeng Liu and Mat Page for useful suggestions and discussions.  
RS carried out part of this work while visiting Tsinghua University in Beijing.


\begin{thebibliography}{99}
 \bibitem[Afonso et al.(2005)]{afo05} Afonso, J., Georgakakis, A., Almeida, C., Hopkins, A. M., 
   Cram, L. E., Mobasher, B., \& Sullivan, M. 2005, ApJ, 624, 135
 \bibitem[Arnaud(1996)]{arn96} Arnaud, K. A. 1996, Astronomical Data Analysis 
   Software and Systems V, eds. G. Jacoby and J. Barnes, ASP Conf. Series volume 101, 17
 \bibitem[Asai et al.(1996)]{asa96} Asai, K., Dotani, T., Mitsuda, K., Hoshi, R., Vaughan, B., 
     \& Tanaka, Y. 1996, PASJ, 48, 257
 \bibitem[Cackett et al.(2005)]{cac05} Cackett, E. M., et al. 2005, ApJ, 620, 922
 \bibitem[Campana, Mereghetti \& Sidoli(1997)]{cam97} Campana, S., Mereghetti, S., 
   \& Sidoli, L. 1997, A\&A, 320, 783
 \bibitem[Campana et al.(1998)]{cam98} Campana, S., Colpi, M., Mereghetti, S., Stella, L., 
   \& Tavani, M. 1998, A\&ARv, 8, 279
 \bibitem[Campana \& Stella(2003)]{cam03} Campana, S., \& Stella, L. 2003, ApJ, 597, 474
 \bibitem[Campana et al.(2004)]{cam04} Campana, S., Israel, G. L., Stella, L., Gastaldello, F., 
    \& Mereghetti, S. 2004, ApJ, 601, 474
 \bibitem[Evans et al.(2007)]{eva07} Evans, P. A., et al. 2007, A\&A, 469, 379
 \bibitem[Evans et al.(2009)]{eva09} Evans, P. A., et al. 2009, MNRAS, 397, 1177
 \bibitem[Farrell et al.(2009)]{far09} Farrell, S. A., Webb, N. A., Barret, D., 
   Godet, O., \& Rodrigues, J. M. 2009, Nature, 460, 73
 \bibitem[Frank, King \& Raine(2002)]{fra02} Frank, J., King, A., 
   \& Raine,  D. J. 2002, Accretion Power in Astrophysics, Cambridge Univ. Press,   Cambridge  
 \bibitem[Godet et al.(2009)]{god09} Godet, O., Barret, D., Webb, N. A., Farrell, S. A., 
   \& Gehrels, N. 2009, ApJ, 705, L109
\bibitem[Heinke et al.(2003a)]{hei03a} Heinke, C. O., Grindlay, J. E., Edmonds, P. D., Lloyd, D. A., 
     Murray, S. S., Cohn, H. N., \& Lugger, P. M. 2003, ApJ, 590, 809
\bibitem[Heinke et al.(2003b)]{hei03b} Heinke, C. O., Grindlay, J. E., Lugger, P. M., Cohn, H. N., 
     Edmonds, P. D., Lloyd, D. A., \& Cool, A. M. 2003, ApJ, 598, 501
\bibitem[Heinke et al.(2005)]{hei05} Heinke, C. O., Grindlay, J. E., \& Edmonds, P. D. 2005, ApJ, 622, 556
\bibitem[James et al.(2000)]{jam00} James, D. J., Jardine, M. M., Jeffries, R. D., Randich, S., 
    Collier Cameron, A., \& Ferreira, M. 2000, MNRAS, 318, 1217
 \bibitem[Jonker et al.(2004)]{jon04} Jonker, P. G., Galloway, D. K., McClintock, J. E., 
   Buxton, M., Garcia, M., \& Murray, S. 2004, MNRAS, 354, 666
 \bibitem[Jonker et al.(2005)]{jon05} Jonker, P. G., Campana, S., Steeghs, D., Torres, M. A. P., 
    Galloway, D. K., Markwardt, C. B., Chakrabarty, D., \& Swank, J. 2005, MNRAS, 361, 511
 \bibitem[Jonker et al.(2007)]{jon07} Jonker, P. G., Steeghs, D., Chakrabarty, D., \& 
     Juett, A. M. 2007, ApJ, 665, L147 
 \bibitem[Makishima et al.(1986)]{mak86} Makishima, K., Maejima, Y.,
        Mitsuda, K., Bradt, H. V., Remillard, R. A., Tuohy, I. R.,
        Hoshi, R., \& Nakagawa, M. 1986, ApJ, 308, 635
 \bibitem[Rutledge et al.(2002)]{rut02} Rutledge, R. E., Bildsten, L., Brown, E. F., Pavlov, G. G., 
       \& Zavlin, V. E. 2002, ApJ, 577, 346
\bibitem[Schmitt, Fleming \& Giampapa(1995)]{sch95} Schmitt, J. H. M. M., Fleming, T. A., 
    \& Giampapa, M. S. 1995, ApJ, 450, 392
 \bibitem[Soria et al.(2010)]{sor10} Soria, R., Hau, G. K. T., Graham, A. W., Kong, A. K. H., 
   Kuin, N. P. M., Li, I.-H., Liu, J.-F., \& Wu, K. 2010, MNRAS, in press
 \bibitem[Webb et al.(2010)]{web10} Webb, N. A., Barret, D., Godet, O., Servillat, M., 
   Farrell, S. A., \& Oates, S. R. 2010, ApJ, 712, L107
 \bibitem[Wiersema et al.(2010)]{far10} Wiersema, K., Farrell, S. A., Webb, N., Servillat, M., 
	Maccarone, T. J., Barret, D., \& Godet, O. 2010, ApJ Letters, in press
 \bibitem[Zampieri et al.(1995)]{zam95} Zampieri, L., Turolla, R., Zane, S., 
   \& Treves, A. 1995, ApJ, 439, 849 
 \bibitem[Zavlin, Pavlov \& Shibanov(1996)]{zav96} Zavlin, V. E., Pavlov, G. G., 
   \& Shibanov, Y. A. 1996, A\&A, 315, 141

\end{thebibliography}
\end{document}